\documentclass[referee]{raa}            

\usepackage{graphicx,times}             
\usepackage{natbib}
\usepackage{epstopdf}
\usepackage{amssymb,amsmath}
\usepackage{subcaption}
\bibpunct{(}{)}{;}{a}{}{,}

\usepackage[pagebackref=true]{hyperref}

\begin{document}

  \title{The Impact of ${}^{12} \mathrm{C}(\alpha, \gamma)^{16} \mathrm{O}$ Reaction on the Evolution of He Stars}

   \volnopage{Vol.0 (20xx) No.0, 000--000}      
   \setcounter{page}{1}          

   \author{Gang Long 
      \inst{1,2,3}
   \and Yu Wang
   \inst{1,2,3}      
   \and Dongdong Liu
      \inst{1,2,3}
   \and Jianguo Wang
    \inst{1,2,3}  
   \and Bo Wang
   \inst{1,2,3}  
   }

   \institute{Yunnan Observatories, Chinese Academy of Sciences,
             Kunming 650216, China; {\it liudongdong@ynao.ac.cn; wangjg@ynao.ac.cn; wangbo@ynao.ac.cn}\\
        \and
            International Centre of Supernovae, Yunnan Key Laboratory, Kunming 650216, China\\
        \and
             University of Chinese Academy of Sciences, Beijing 100049, China\\
\vs\no
   {\small Received 20xx month day; accepted 20xx month day}}

\abstract{ 
	The ${}^{12} \mathrm{C}(\alpha, \gamma)^{16} \mathrm{O}$ reaction is one of the most important reactions in the evolution of massive stars,  yet its rate is still highly uncertain. In this work,
	we investigated how variations in the ${}^{12} \mathrm{C}(\alpha, \gamma)^{16} \mathrm{O}$ reaction rate affect the evolution of a 14 $\rm M_{\odot}$ He star using the MESA code. Our simulations indicate that the ${}^{12} \mathrm{C}(\alpha, \gamma)^{16} \mathrm{O}$ reaction rate determines the conditions for C burning, affecting its explodability.
    As the reaction rate increases, central C-burning becomes neutrino-dominated, transitioning from the convective to the radiative regime. This leads to higher compactness and a larger iron core,  indicating a more compact pre-SN core structure that is difficult to explode. Conversely, lower reaction rates shorten the C-burning lifetime and trigger earlier central Ne ignition, which counteracts core contraction. This results in reduced compactness and iron core mass. 
    We also found that variations in reaction rates shift the location of the last C-burning shell. When this shell exceeds the mass coordinate used for compactness evaluation, the overall compactness increases significantly. Although the Si- and O-burning convective shells decrease compactness, the overall increase remains unaffected. This work suggests that the ${}^{12} \mathrm{C}(\alpha, \gamma)^{16} \mathrm{O}$ reaction play an important role in the pre-SN core structure and potentially impact the explodability of massive He stars.
}

   \authorrunning{Gang long et al. }            
   \titlerunning{The Impact of ${}^{12} \mathrm{C}(\alpha, \gamma)^{16} \mathrm{O}$ Reaction on the Evolution of He Stars}  

   \maketitle

%
%
\section{Introduction}           

Massive He stars are thought to be stripped products of more massive stars that have lost their hydrogen-rich envelopes through stellar winds or binary interactions \citep[e.g.][]{podsiadlowski1992presupernova,pols2002helium,crowther2007physical,langer2012presupernova,yoon2017towards,dessart2020supernovae}. These stars end their lives as core-collapse supernovae (CCSNe), leaving behind neutron stars (NSs) or black holes (BHs)\citep[e.g.][]{woosley2019evolution,ertl2020explosion}. Their final fates are mainly determined by the carbon-oxygen (CO) core mass $M_{\rm CO}$ and the central carbon mass fraction $X_{\rm i}({}^{12}\rm C)$ at the end of core He burning, both of which are sensitive to the ${}^{12} \mathrm{C}(\alpha, \gamma)^{16} \mathrm{O}$ reaction \citep{Weaver1993PhR,Woosley2007PhR,Tur2007ApJ,Tur2010ApJ,chieffi2020presupernova,patton2020towards,sukhbold2020missing,farmer2020constraints,schneider2021pre,schneider2023bimodal,laplace2025s}. However, its reaction rate remains highly uncertain \citep{deboer2017c}. 
Therefore, it is important to investigate the influence of the uncertainty in the ${}^{12} \mathrm{C}(\alpha, \gamma)^{16} \mathrm{O}$ reaction rate on the evolution and the pre-supernova (pre-SN) core structure.

The ${}^{12} \mathrm{C}(\alpha, \gamma)^{16} \mathrm{O}$ reaction is one of the most important reactions in the evolution of massive stars \citep{Weaver1993PhR,woosley2003nuclear}. Its rate determines the ratio of  ${}^{12} \mathrm{C} $ and ${}^{16} \mathrm{O}$ after He burning by competing with the well-determined triple-alpha ($3\alpha$) reaction \citep{Buchmann1996ApJ,Imbriani2001reaction,Kunz2002ApJ,woosley2003nuclear,Eid2004Evolution,Tur2007ApJ,Woosley2007PhR,Austin2014PhRvL}. Variations in these abundances have significant effects on subsequent burning stages, as well as the core structure and nucleosynthesis  \citep{Imbriani2001reaction,Tur2007ApJ,Tur2010ApJ,west2013ApJ,farmer2020constraints}. 
For instance, \cite{Imbriani2001reaction} investigated how variations in the ${}^{12} \mathrm{C}(\alpha, \gamma)^{16} \mathrm{O}$ reaction impact the residual ${}^{12}$C mass fraction after helium burning, which determines whether C burning proceeds radiatively or convectively, thereby affecting the pre-SN core structure and the explodability of stars \citep{sukhbold2014compactness,sukhbold2016core,sukhbold2020missing,patton2020towards,schneider2021pre}. 
Additionally, several studies have explored the impact of uncertainties in this reaction on nucleosynthesis \citep{Weaver1993PhR,Tur2010ApJ,Farmer2023ApJ}.
 Notably, uncertainties in this reaction rate also affect the ranges of both the pair-instability (PI) mass gap and the BH mass gap \citep{Farmer2019ApJ,farmer2020constraints,Costa2021MNRAS,Woosley2021ApJ}.
Motivated by these studies, we focus on the impact of the ${}^{12} \mathrm{C}(\alpha, \gamma)^{16} \mathrm{O}$ reaction rate on the pre-SN core structure and the explodability of the star.

Previous studies investigated the connection between the final pre-SN structure and the outcome of neutrino-driven CCSN explosions, aiming to predict the explodability of progenitors, i.e., whether the progenitors will explode as a SN or not \citep[e.g.][]{o2011black,ugliano2012progenitor,ertl2016two,muller2016simple}. \cite{o2011black} proposed the compactness parameter $\xi_{2.5}$ to characterize the core structure and proposed it as a criterion to assess stellar explodability. Progenitors with higher compactness are more difficult to explode  \citep[e.g.][]{o2011black,ugliano2012progenitor,sukhbold2014compactness,muller2016simple,sukhbold2016core,sukhbold2020missing,chieffi2020presupernova}. Additionally, a two-parameter criterion related to the density and entropy at the Si/O interface was used to distinguish between successful and failed explosions \citep{ertl2016two,ertl2020explosion}. 
Thus, we use these parameters to characterize the final core structure and explore the explodability of progenitors.

In this work, we aim to investigate the impact of uncertainties in the ${}^{12} \mathrm{C}(\alpha, \gamma)^{16} \mathrm{O}$ reaction rate on the evolution of massive He stars and predict their final fates. 
The paper is organized as follows. 
In Section \ref{sect:models}, we describe the basic assumptions and numerical methods. The simulation results are presented in Section \ref{sect:Results}. Finally, we discuss our results in Section \ref{sect:Discussion}, and conclude in Section \ref{sect:Conclusion}.

\section{method}\label{sect:models}
To investigate the impact of the ${}^{12}\mathrm{C}(\alpha, \gamma)^{16}\mathrm{O}$ reaction rate on the core structure of massive He stars, we use the MESA code \citep[version 10398;][]{paxton2011modules,paxton2013modules,paxton2015modules,paxton2018modules,paxton2019modules} to simulate the evolution of the 14 $\rm M_{\odot}$ He star models with different ${}^{12}\mathrm{C}(\alpha, \gamma)^{16}\mathrm{O}$ reaction rates, from the He zero-age main sequence (He-ZAMS) to the onset of core collapse, defined as when the contraction velocity in the iron core boundary reaches $1000 \, \mathrm{km} \, \mathrm{s}^{-1}$. 
The initial He star models are constructed with a metallicity of Z = 0.02 \citep[see also][]{aguilera2022stripped}, and we adopt the wind mass-loss prescription from \citet{yoon2017towards}. 
For the ${}^{12}\mathrm{C}(\alpha, \gamma)^{16}\mathrm{O}$ reaction rates, we use MESA’s default rates from \citet[][hereafter NACRE]{angulo1999compilation}, and also consider two widely used compilations from \citet[hereafter CF88]{caughlan1988thermonuclear} and \citet[hereafter Kunz]{Kunz2002ApJ}, both of which report lower reaction rates compared to the MESA default.
Convection is treated using the standard mixing-length theory \citep{bohm1958}, with a mixing-length parameter of $\alpha_{\rm MLT} = 2.0$, and convective boundaries are determined using the Ledoux criterion. 
We apply exponential overshooting with a parameter of $f_{\rm ov} = 0.016$ during core He burning \citep{Herwig2000A&A}. Additionally, we adopt a semiconvection efficiency parameter of $\alpha_{\rm SEM} = 1.0$, and a thermohaline diffusion coefficient of $D_{\rm TH} = 1.0$.

To evaluate the explodability of progenitors, we focus on the compactness criterion and the two-parameter criterion, due to their widespread use in predicting whether iron core collapse will result in a successful or failed neutrino-driven SN mechanism.
The compactness parameter $\xi_{2.5}$ characterizes the core structure and is defined as \citep{o2011black}
\begin{equation}\label{xi}
	\xi_M \equiv \frac{M / \rm M_{\odot}}{R(M) / 1000 \mathrm{~km}}, 
\end{equation}
where $M$ and $R(M)$ are the enclosed mass and the radius as a function of the mass coordinate, respectively. The typical mass used to evaluate explodability is $M = 2.5\, \rm M_{\odot}$. The parameter $\xi_{2.5}$
is closely related to the iron core mass, with both quantities commonly used to characterize the final core structure \citep[e.g.][]{brown2001formation,o2011black,sukhbold2014compactness,ertl2016two,muller2016simple,patton2020towards,laplace2025s}. The two-parameter criterion consists of $M_{4}$ and $\mu_{4}$ proposed by \cite{ertl2016two}:
\begin{equation}\label{mu_4}
	\left.\mu_4 \equiv \frac{d m / \rm M_{\odot}}{d r / 1000 \mathrm{~km}}\right|_{s=4}.
\end{equation}
Here, $M_{4}$ and $\mu_{4}$ represent the mass and radial coordinates where the entropy per baryon reaches $4.0\, k_{\rm B}$, and they are related to the density and entropy distributions. Meanwhile, $\mu_{4}$ and $\mu_{4} \times M_{4}$ 
are indirectly related to the mass accretion rate at the time of explosion and the accretion luminosity, respectively.
Based on this, the likely explosion and collapse models are distinguished by the relation $\mu_4 = 0.294 \mu_4 M_4 + 0.0468$ from \cite{ertl2020explosion}.
Additionally, we use the semianalytic approach of \cite{muller2016simple}, which is based on the neutrino-driven CCSN explosions, to predict whether stars explode as SNe or collapse into BHs obtain the properties of successful SN explosion, such as explosion energy $E_{\rm exp}$ and gravitational remnant mass $M_{\rm rm,grav}$.

\begin{table}
	\caption{Pre-SN Properties with Varying ${}^{12}\mathrm{C}(\alpha,\gamma){}^{16}\mathrm{O}$ Reaction Rates}\label{physical parameters}
	\begin{center}
		\begin{tabular}{
				l@{\hspace{1.29mm}}c@{\hspace{1.29mm}}c@{\hspace{1.29mm}}
				c@{\hspace{1.29mm}}c@{\hspace{1.29mm}}c@{\hspace{1.29mm}}
				c@{\hspace{1.29mm}}c@{\hspace{1.29mm}}c@{\hspace{1.29mm}}
				c@{\hspace{1.29mm}}c@{\hspace{1.29mm}}c@{\hspace{1.29mm}}
				c@{\hspace{1.29mm}}c@{\hspace{1.29mm}}c@{\hspace{1.29mm}}
				c@{\hspace{1.29mm}}c@{\hspace{1.29mm}}r@{\hspace{1.29mm}}
			}
			\hline\hline
			
			$\rm Rate$ & $M_{\rm CO}$ &  $X_{\rm i}({}^{12}\mathrm{C})$ & $X_{\rm i}({}^{16}\mathrm{O})$ & $M_{\text{C-free}}$  & $\xi_{2.5}$ &$\mu_{4}$&$M_{4}$& $s_{\text{c}}$  & $M_{\rm f}$ & $M_{\rm Fe}$ &  log$(-E_{\text{bind}})$ & $E_{\text{exp}}$ & $\text{Fate}$  &$M_{\text{rm,grav}}$
			\\
			& $\left[\mathrm{M_{\odot}}\right]$ &  &  &   $\left[\mathrm{M_{\odot}}\right]$  &    &  &  &$\left[N_{\mathrm{A}} k_{\mathrm{B}}\right] $ &$\left[\mathrm{M_{\odot}}\right]$&$\left[\mathrm{M_{\odot}}\right]$ &$\left[\text{erg}\right] $ & $\left[10^{51}\text{erg}\right] $  & &$\left[\mathrm{M_{\odot}}\right]$ \\
			\hline
			\text{NACRE} & 7.12 & 0.260&0.713 & 2.73 &  0.44 &  0.13 & 2.15 &   1.06 & 8.29 &  1.79 &   51.61& $\cdots$ & BH & 8.29 \\
			\text{CF88} &  7.18 & 0.286&0.687 & 2.06 &  0.16 &  0.07 & 1.77 &   0.92  & 8.34 &  1.54 &   51.40&0.57 & NS & 1.55 \\		
			\text{Kunz} &  7.21 & 0.298& 0.675& 1.75 &  0.08 &  0.03 & 1.52 &   0.79  & 8.36 &  1.46 &   51.22&0.23& NS & 1.34 \\		    
			
			\hline
		\end{tabular}
	\end{center}
	{\bf Notes:} The table summarizes key pre-SN properties for the 14 $\rm M_{\odot}$ He star with varying ${}^{12}\mathrm{C}(\alpha,\gamma){}^{16}\mathrm{O}$ reaction rates. Three sets of reaction rates from NACRE \citep{angulo1999compilation}, CF88 \citep{caughlan1988thermonuclear}, and Kunz \citep{Kunz2002ApJ} were employed, with the rates decreasing sequentially.
   The following quantities are provided: the CO core mass $M_{\rm CO}$, the central ${}^{12}\mathrm{C}$ mass fraction $X_{\rm i}({}^{12}\mathrm{C})$, and the central ${}^{16}\mathrm{O}$ mass fraction $X_{\rm i}({}^{16}\mathrm{O})$, all measured at the end of core He burning when the central helium abundance drops below $10^{-4}$; the C-free core mass $M_{\text{C-free}}$, defined as the mass coordinate where the $X_{\rm i}({}^{12}\mathrm{C})$ falls below $10^{-5}$; the compactness parameter $\xi_{2.5}$; the central specific entropy at core collapse $s_{\rm c}$; the radial mass derivative $\mu_{4}$ and enclosed mass $M_{4}$ at a specific entropy of $s=4$; the final mass $M_{\rm f}$; the iron core mass at core collapse $M_{\rm Fe}$; the gravitational binding energy of the material above the iron core $-E_{\rm bind}$; the explosion energy $E_{\rm exp}$; the predicted final fate; and the gravitational remnant mass $M_{\rm rm,grav}$. Both $E_{\rm exp}$ and $M_{\rm rm,grav}$ are computed following the semi-analytic approach of \cite{muller2016simple}.
	
\end{table}

\section{Pre-SN evolution and core structure}\label{sect:Results}

In this section, we examine the impact of the uncertainty in the ${}^{12}\mathrm{C}(\alpha,\gamma){}^{16}\mathrm{O}$ reaction rate on the evolution of massive He stars, particularly their chemical composition and core structure. 
As ${}^{4}\mathrm{He}$ is depleted and ${}^{12}\mathrm{C}$ accumulates during core He burning, the ${}^{12}\mathrm{C}(\alpha,\gamma){}^{16}\mathrm{O}$ reaction becomes dominant and strongly affects the subsequent evolution.
To further illustrate this, we explore its effect on the mass fractions of ${}^{12}\mathrm{C}$ and the CO core mass during core He burning, as well as their influence on subsequent carbon and advanced burning phases, particularly the evolution of compactness and core structure.

\subsection{Core composition after core helium burning}\label{sect:He-burning phase}

\begin{figure} 
	\centering
	\includegraphics[width=12.0cm, angle=0]{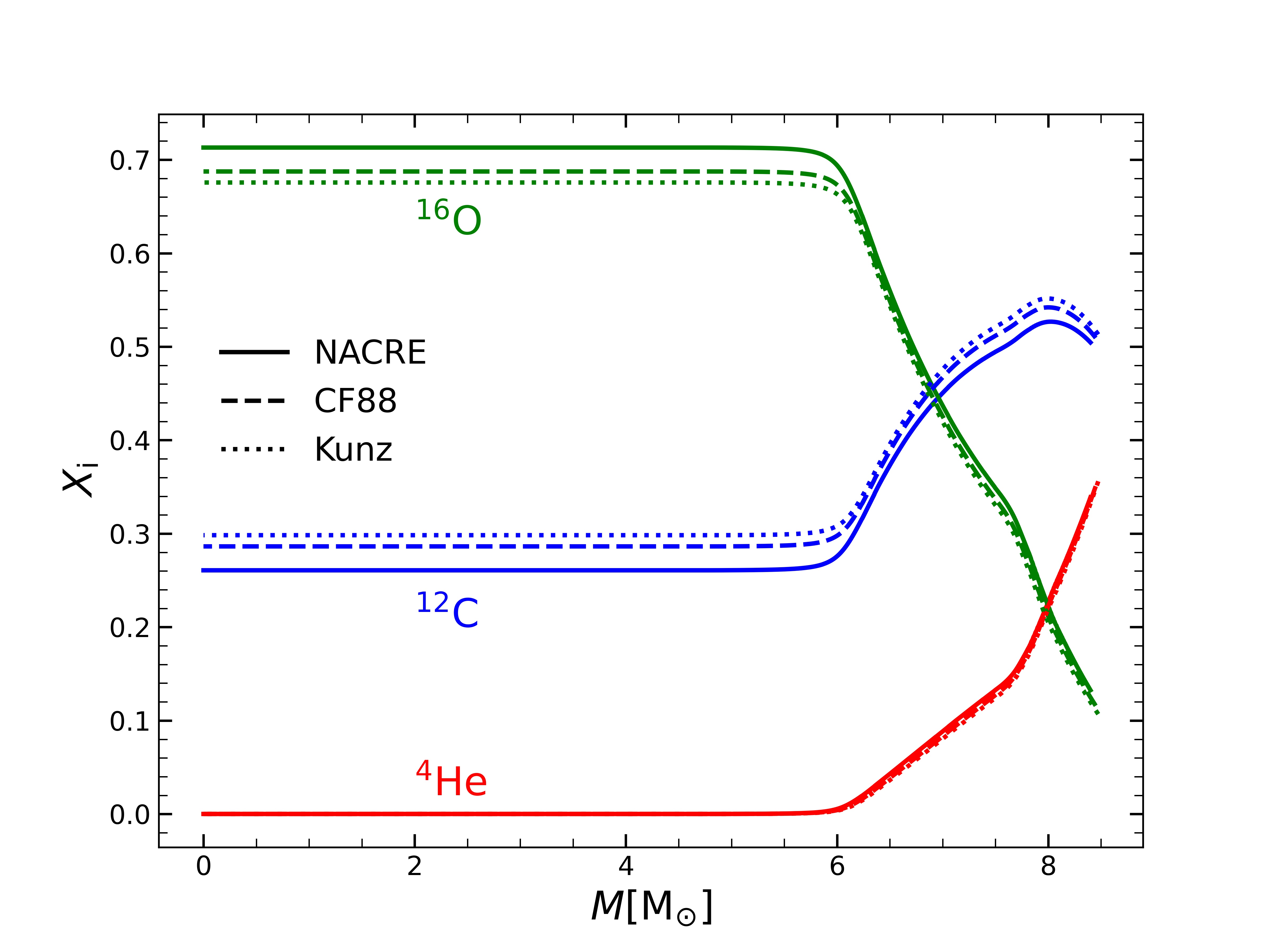}
	\caption{Profiles of abundance of ${}^{4}\mathrm{He}$, ${}^{12}\mathrm{C}$, and ${}^{16}\mathrm{O}$ at the end of core He burning for 14 $\rm M_{\odot}$ He stars, calculated using different reaction rates.	
    The solid, dashed, and dotted lines correspond to reaction rates from NACRE, CF88, and Kunz, respectively. The red, blue, and green lines represent the abundances of ${}^{4}\mathrm{He}$, ${}^{12}\mathrm{C}$, and ${}^{16}\mathrm{O}$, respectively.
		\label{fig:mass_fraction}}
\end{figure}

\begin{figure} 
	\centering
	\includegraphics[width=12.0cm, angle=0]{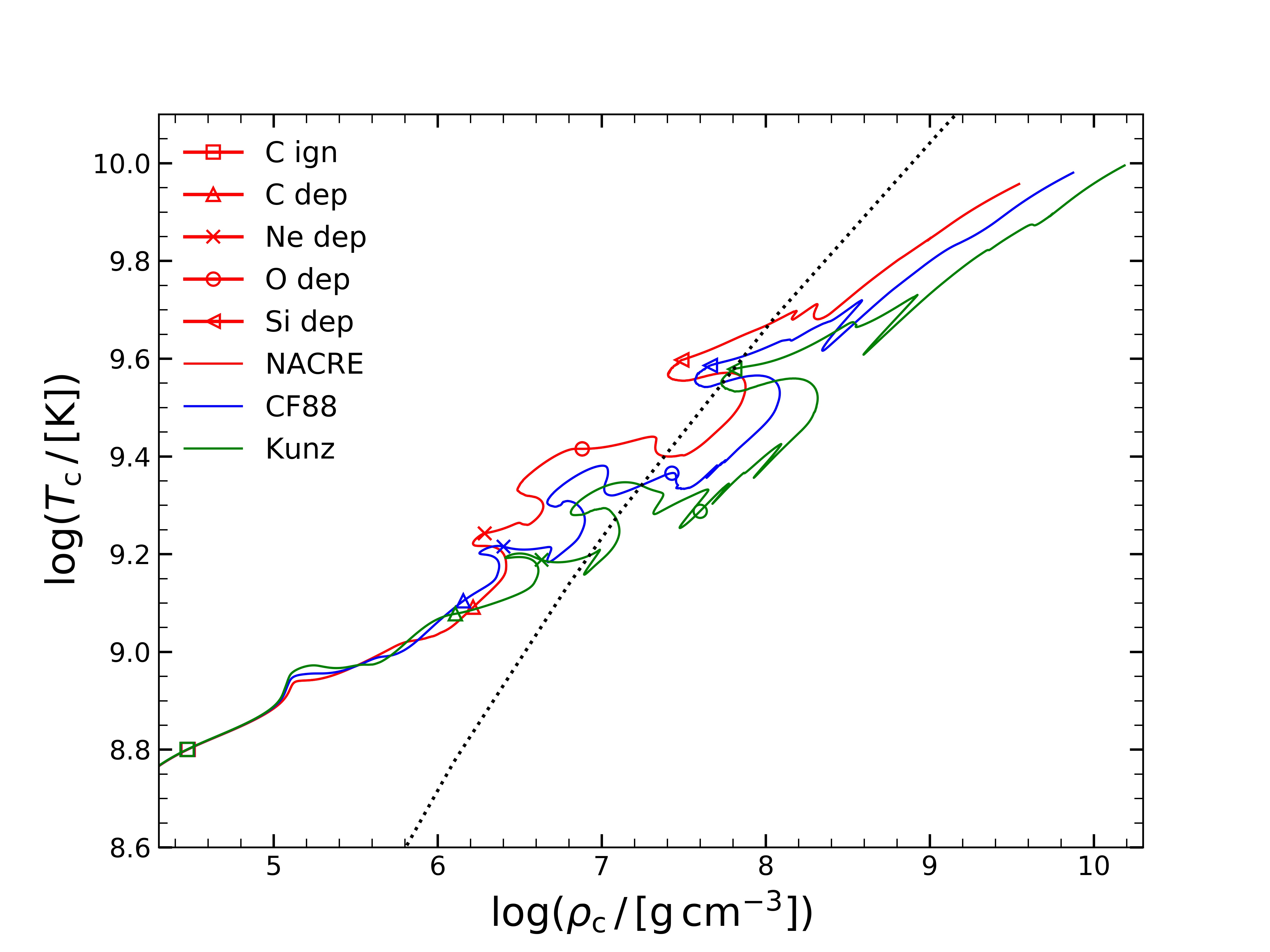}
	\caption{The evolution of the central temperature $T_{\rm c}$ as a function of central density $\rho_{\rm c}$ for the 14 $\rm M_{\odot}$ He stars with three different ${}^{12}\mathrm{C}(\alpha, \gamma)^{16}\mathrm{O}$ reaction rates (see Table \ref{physical parameters}), with each reaction rate represented by a different colored curve.
    Key evolutionary phases are indicated by the following symbols: pentagon for central He depletion, square for central C ignition at $T_{\rm c}=8 \times 10^8 \, \mathrm{K}$, triangle for central C depletion, cross for central Ne depletion, open circle for central O depletion, and left-pointing triangle for central Si depletion. The depletion of these elements is defined as the stage when their central mass fraction $X_{\rm i}$ falls below $10^{-4}$. The gray dotted line marks the boundary between non-degenerate and degenerate electron gas.
	\label{fig:reaction_rate}}
\end{figure}

Varying the ${}^{12}\mathrm{C}(\alpha, \gamma)^{16}\mathrm{O}$ reaction rate affects the chemical composition of the core.
Specifically, a higher reaction rate results in less ${}^{12}\mathrm{C}$ and more ${}^{16}\mathrm{O}$ at the end of core He burning, while a lower reaction rate leads to the opposite due to fewer $\alpha$ captures by ${}^{12}\mathrm{C}$ \citep{deboer2017c}. Fig. \ref{fig:mass_fraction} shows the abundance profiles of ${}^{4}\mathrm{He}$, ${}^{12}\mathrm{C}$, and ${}^{16}\mathrm{O}$ at the end of core He burning for a 14 $\rm M_{\odot}$ He star, under the various reaction rate scenarios considered. Notably, because the Kunz reaction rate is lower than both the NACRE and CF88 rates, it results in a higher $X_{\rm i}({}^{12}\mathrm{C})$ and a lower $X_{\rm i}({}^{16}\mathrm{O})$ (see also Table \ref{physical parameters}).  For instance, we obtain $X_{\rm i}({}^{12}\mathrm{C}) \approx 0.260$, $X_{\rm i}({}^{12}\mathrm{C}) \approx 0.286$, and $X_{\rm i}({}^{12}\mathrm{C}) \approx 0.298$ for using NACRE, CF88, and Kunz reaction rates, respectively.  However, despite these differences in chemical composition, we find that the $M_{\rm CO}$ at the end of core He burning is insensitive to variations in the ${}^{12}\mathrm{C}(\alpha, \gamma)^{16}\mathrm{O}$ reaction rates. For the models using the NACRE, CF88, and Kunz rates, the values of $M_{\rm CO}$ are $7.12\, \rm M_{\odot}$, $7.18\, \rm M_{\odot}$, and $7.21\, \rm M_{\odot}$, respectively.  The small change in the CO core mass is expected, as it is primarily determined by the convective core mass during core He burning, which is mainly influenced by the total energy from the 3$\alpha$ and ${}^{12}\mathrm{C}(\alpha, \gamma)^{16}\mathrm{O}$ reactions. Since the latter contributes relatively little to energy generation, the 3$\alpha$ process dominates \citep{woosley2002evolution}.

\subsection{Carbon burning phase}\label{sect:C-burning phase}

As shown in Sec. \ref{sect:He-burning phase}, the ${}^{12}\mathrm{C}(\alpha, \gamma)^{16}\mathrm{O}$ reaction affects both $M_{\rm CO}$ and $X_{\rm i}({}^{12}\mathrm{C})$. Even small variations in these quantities can have a significant impact on the subsequent core structure and the final fate \citep{sukhbold2014compactness,sukhbold2020missing,patton2020towards,chieffi2020presupernova}. Fig. \ref{fig:reaction_rate} illustrates the evolutionary tracks of central density and temperature after central He depletion, with each curve corresponding to a different adopted reaction rate, revealing notable differences. These differences are driven by the energy generation rate from C burning, $\epsilon_{\mathrm{n}} \sim X_{\rm i}({}^{12}\mathrm{C})^2 \rho T^{23}$, and the energy loss rate through neutrinos, $\epsilon_{v} \sim T^{12} \rho^{-1}$, where $\rho$ and $T$ represent the density and temperature near the centre \citep{sukhbold2020missing}. Both $\epsilon_{\mathrm{n}}$ and $\epsilon_{v}$ depend on $\rho$ and $T$, which are governed by $M_{\rm CO}$ and $X_{\rm i}({}^{12}\mathrm{C})$. 
For example, during core C burning, the NACRE model results in lower temperatures at a given central density compared to the CF88 and Kunz models, but it shows an overall shift toward higher central temperatures after carbon depletion.

\begin{figure} 
	\centering
	\includegraphics[width=12.0cm, angle=0]{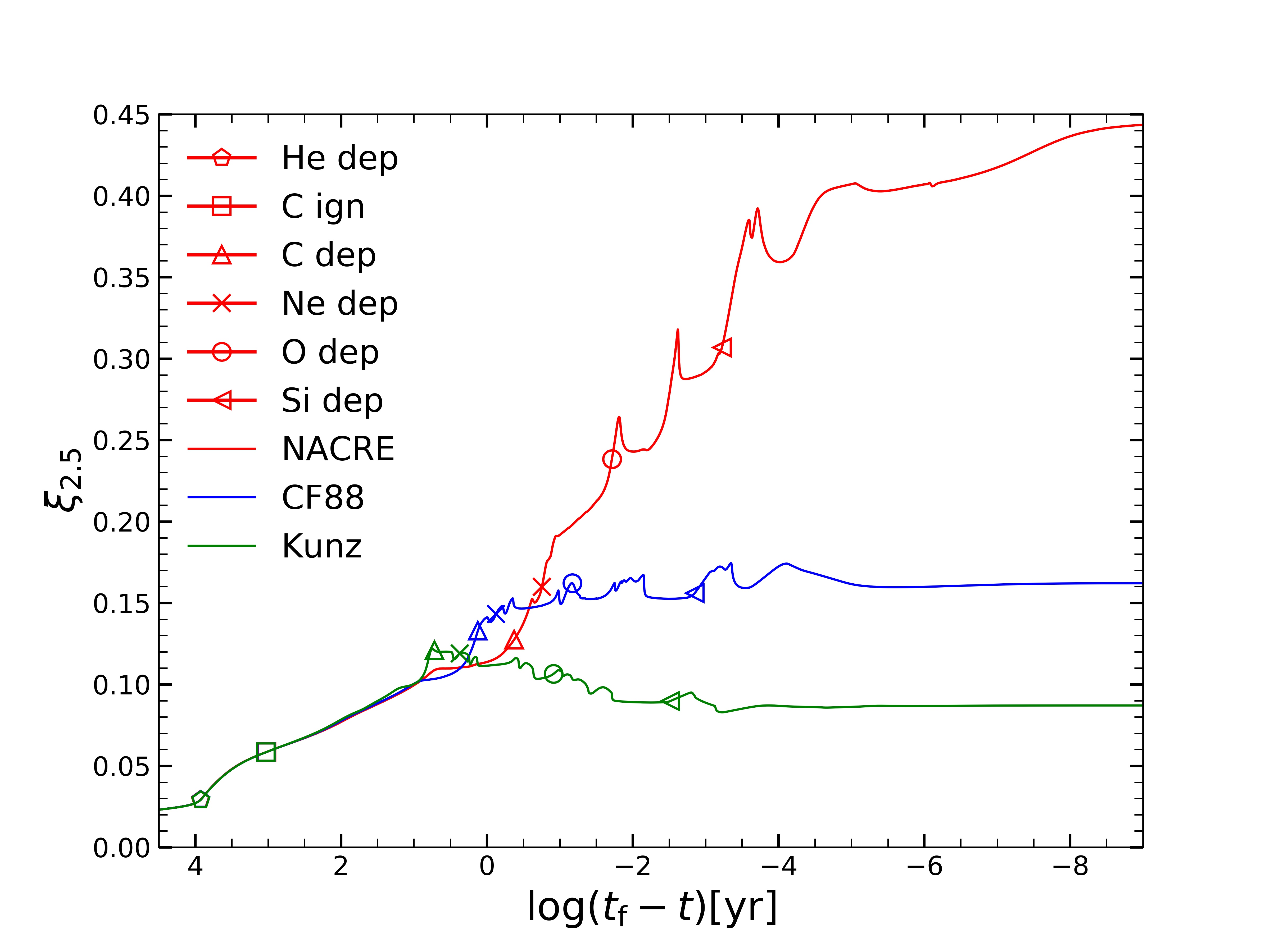}
	\caption{Time evolution of the compactness parameter $\xi_{2.5}$ for three different ${}^{12}\mathrm{C}(\alpha, \gamma)^{16}\mathrm{O}$ reaction rate models, from He-ZAMS to the onset of core collapse. Key evolutionary phases are marked by symbols, as described in Fig. \ref{fig:reaction_rate}
		\label{fig:compactness}}
\end{figure}

\begin{figure}
	\centering
	\begin{minipage}{0.29\textwidth}
		\centering
		\includegraphics[width=1.245\textwidth]{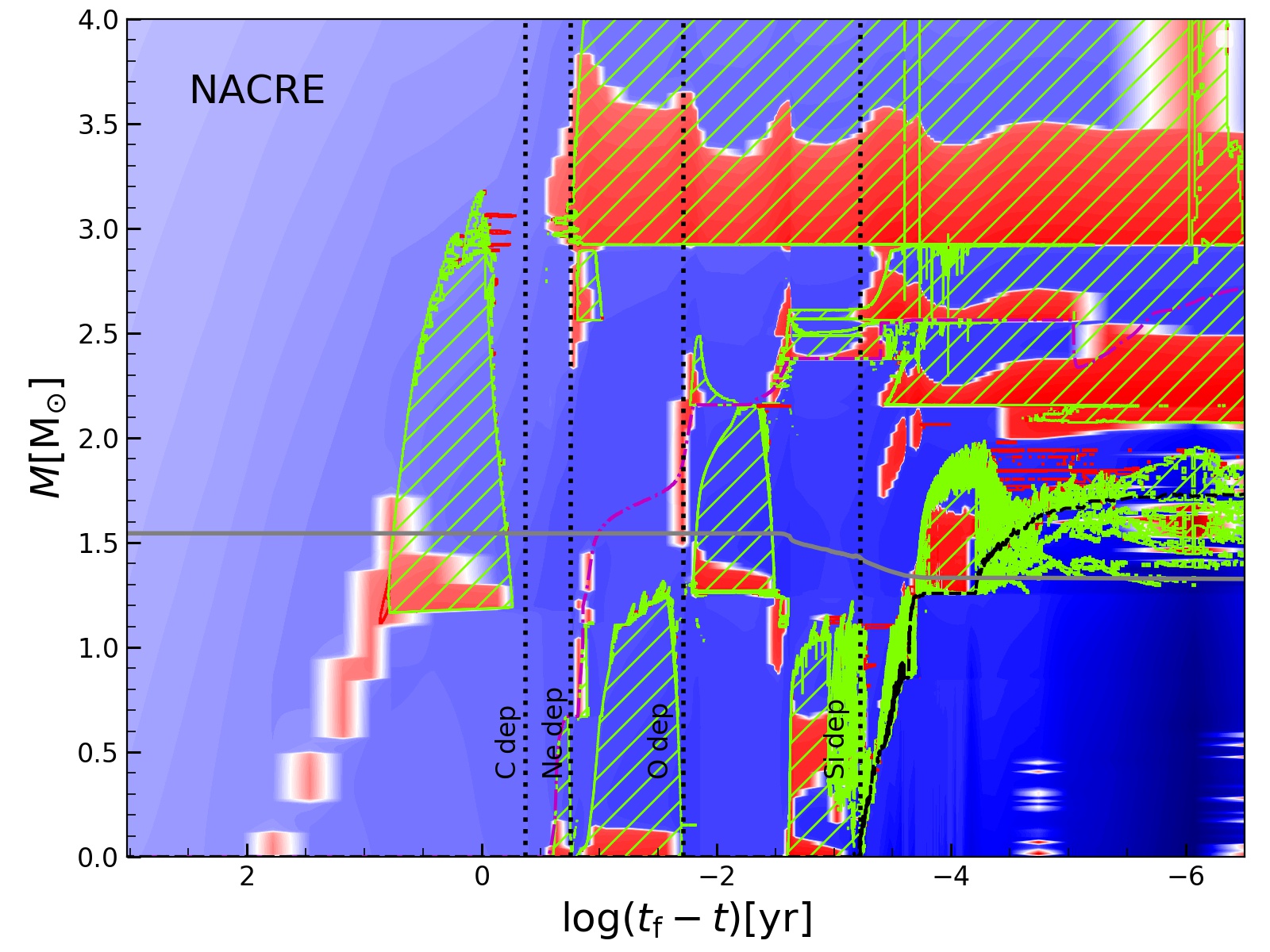}
	\end{minipage}\hfill
	\begin{minipage}{0.29\textwidth}
		\centering
		\includegraphics[width=1.245\textwidth]{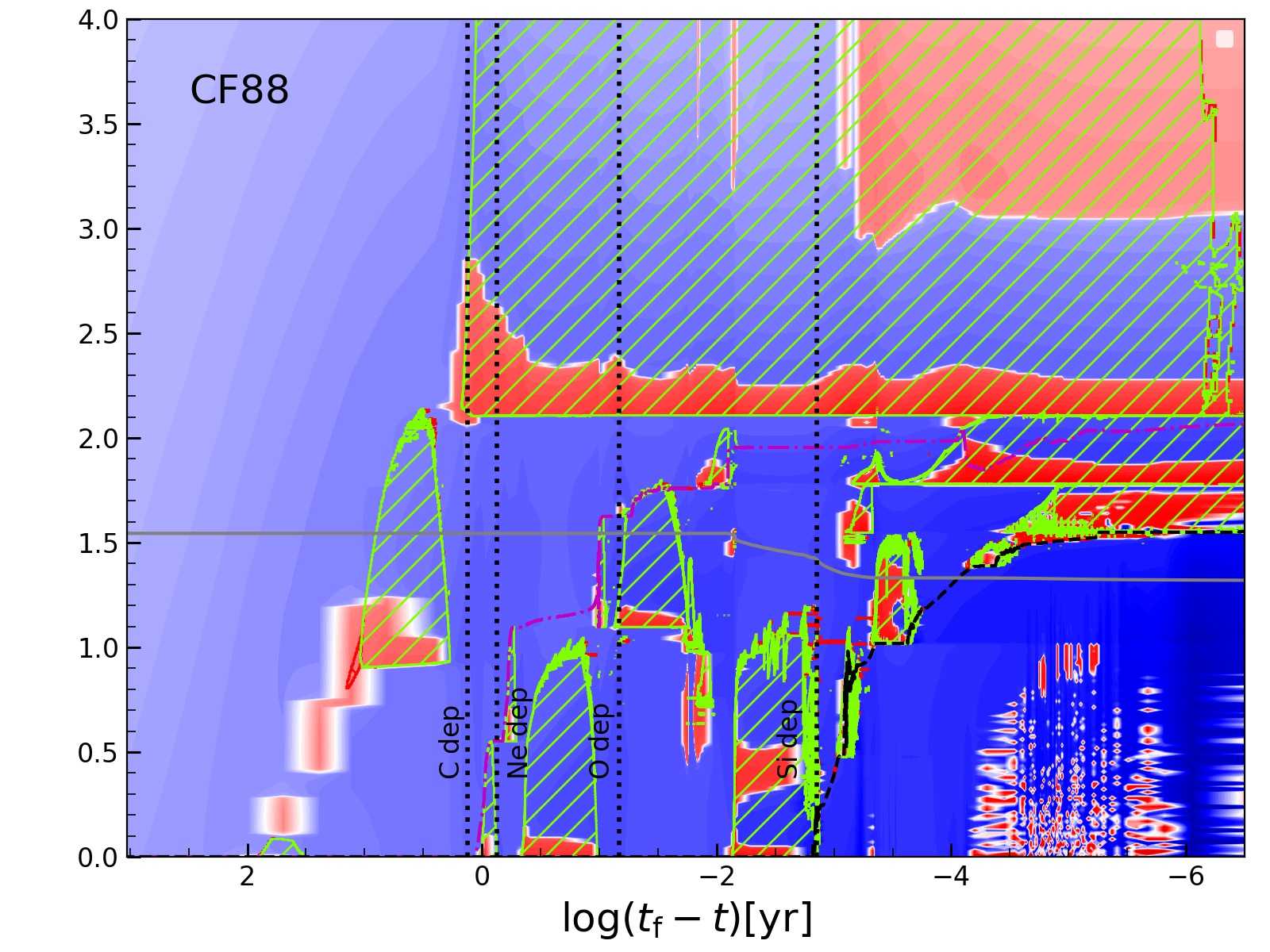}
	\end{minipage}\hfill
	\begin{minipage}{0.29\textwidth}
		\centering
		\includegraphics[width=1.245\textwidth]{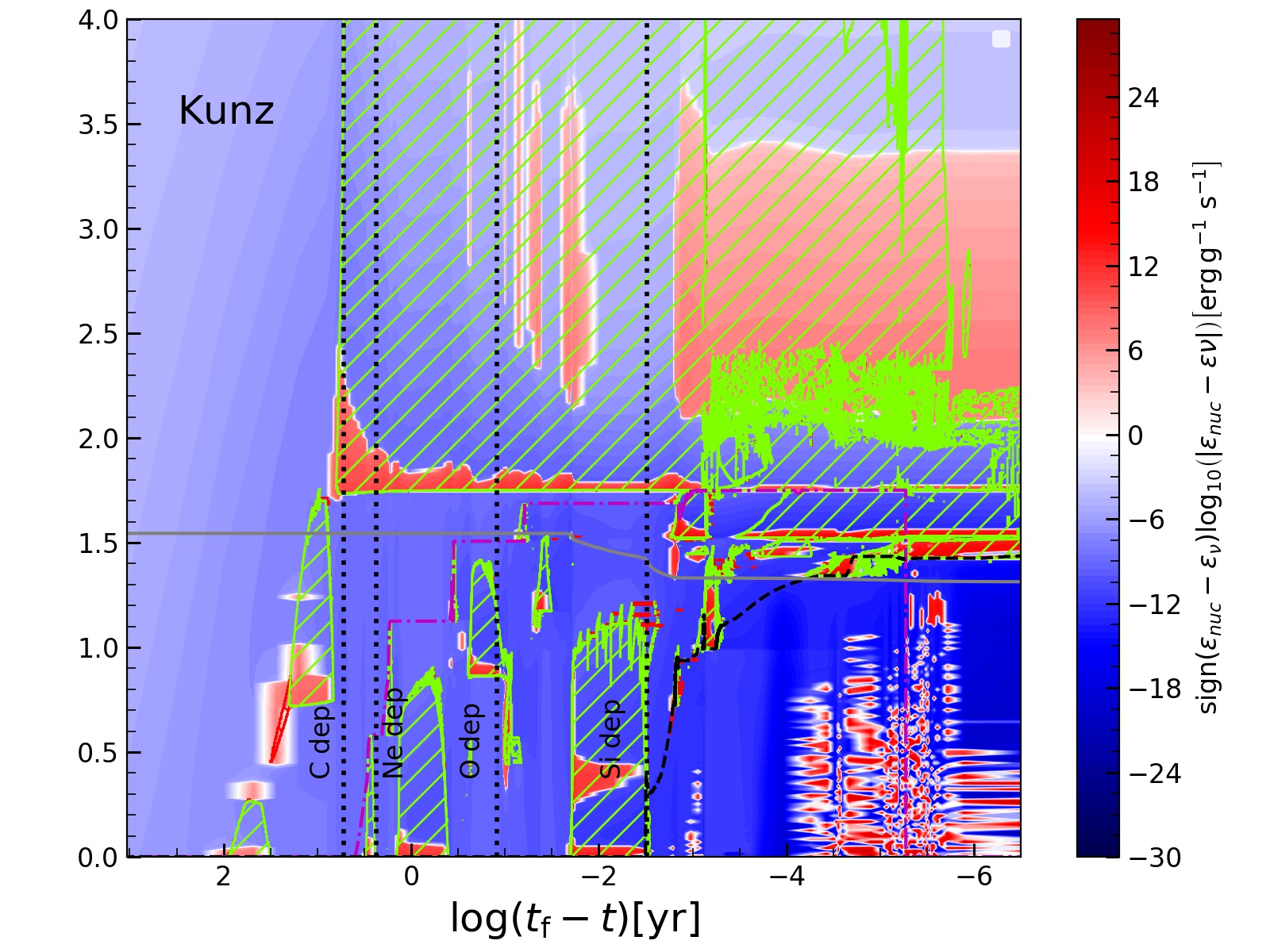}
	\end{minipage}
	\caption{Kippenhahn diagram for a 14 $\rm M_{\odot}$ He stars based on different ${}^{12}\mathrm{C}(\alpha, \gamma)^{16}\mathrm{O}$ reaction rates (NACRE, CF88 and Kunz), from He-ZAMS to the onset of core collapse. The color bar represents the nuclear energy generation rate $\epsilon_{\rm nuc}$ (red) and the neutrino loss rate $\epsilon_{\nu}$ (blue) gradients. The green and red-hatched regions correspond to the convective and semiconvective mixing regions, respectively. The grey solid line represents the classical Chandrasekhar mass $M_{\rm Ch,0}$, as defined by \cite{Timmes1996ApJ}. 
		The burning front is defined as the mass coordinate at which the peak energy generation rate occurs for a specific nuclear burning process, such as the base of the convective burning zone.
		The magenta dash-dotted line represents the C-free core mass $M_{\text{C-free}}$.
   The black dashed line marks the iron core mass, while the dotted lines, from left to right, represent the depletion of central C, Ne, O, and Si.
		\label{fig:reaction_rate_kippen}}
\end{figure}

\begin{figure}
	\centering
	\begin{minipage}{0.29\textwidth}
		\centering
		\includegraphics[width=1.245\textwidth]{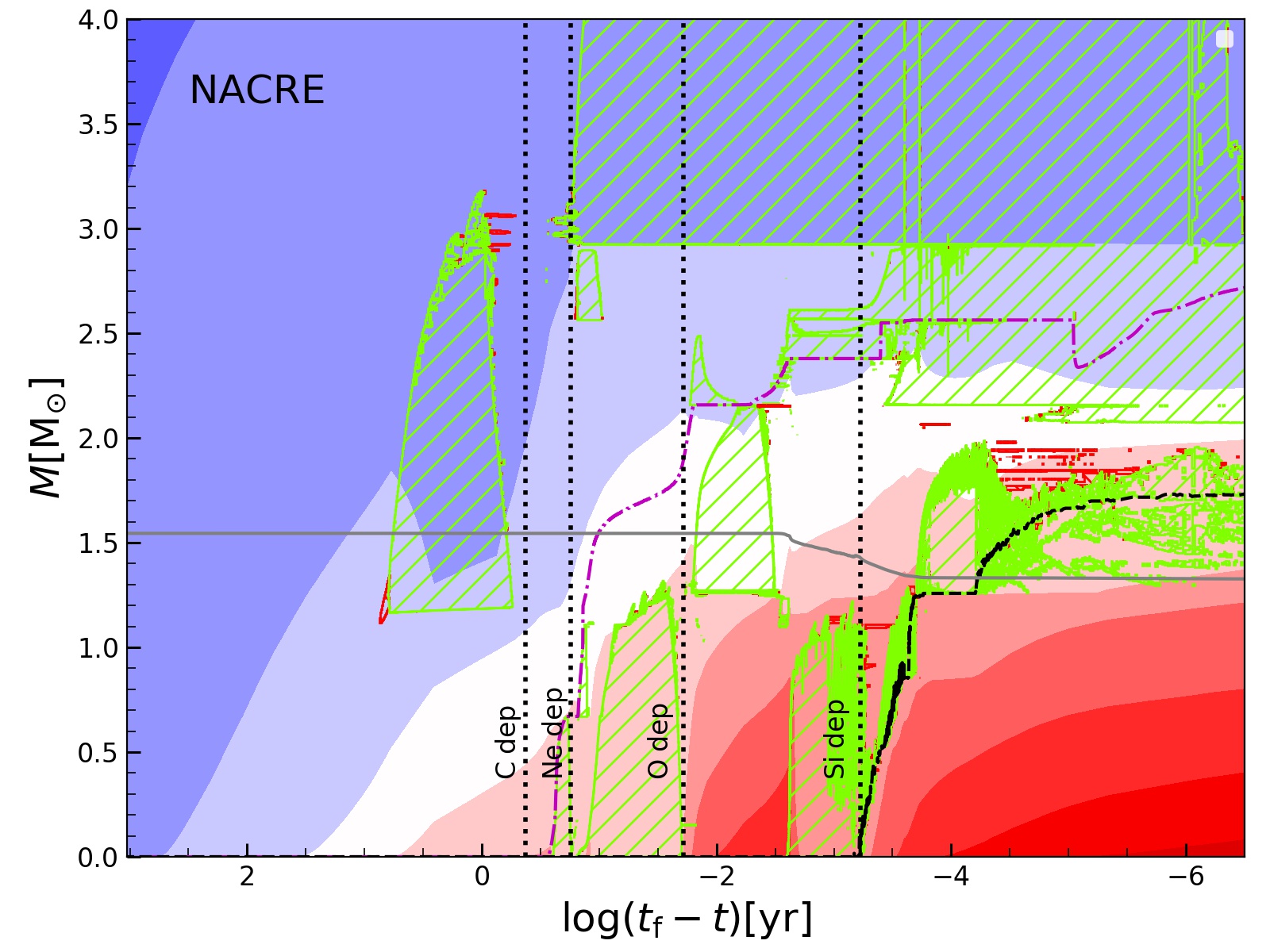}
	\end{minipage}\hfill
	\begin{minipage}{0.29\textwidth}
		\centering
		\includegraphics[width=1.245\textwidth]{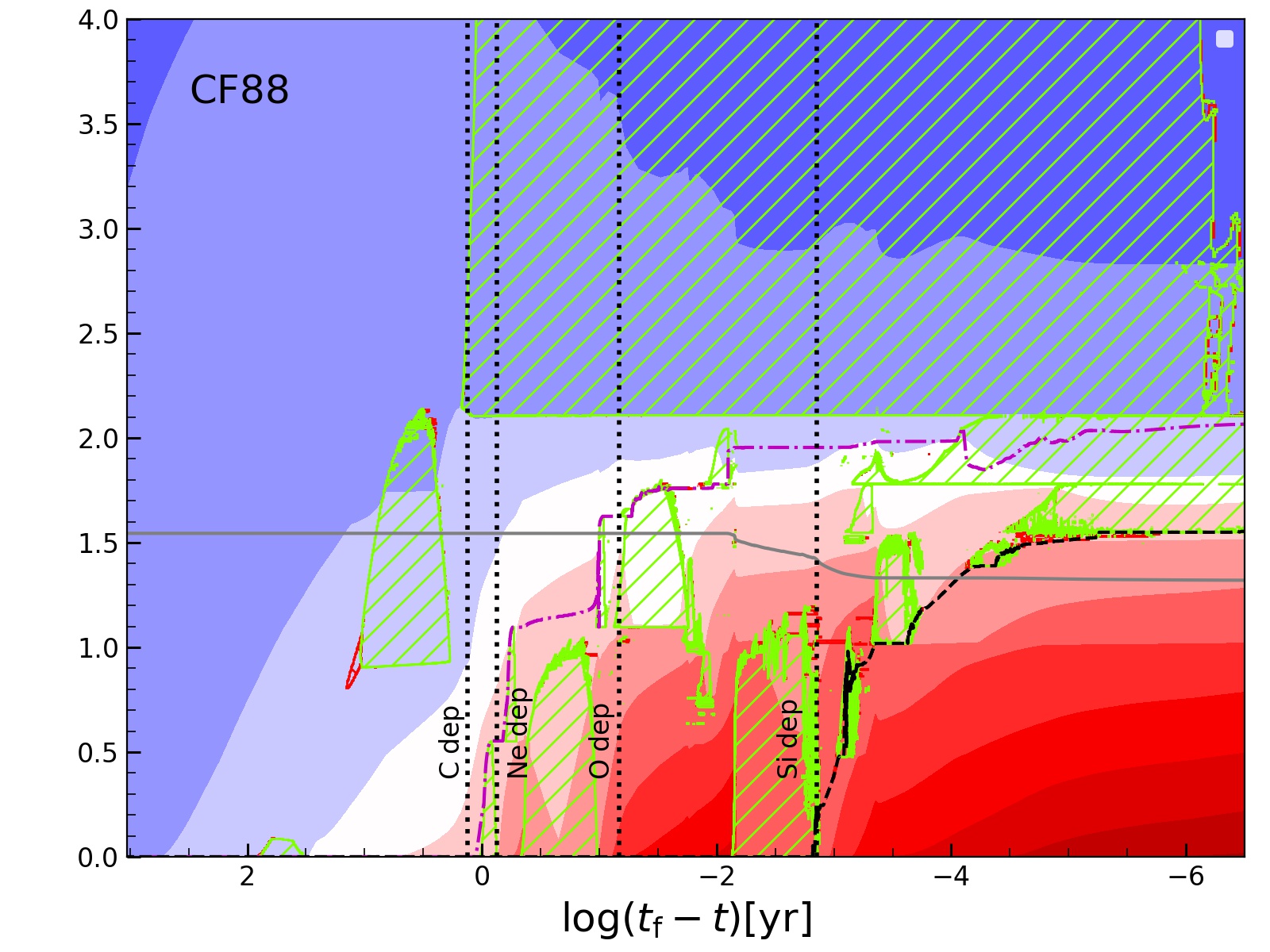}
	\end{minipage}\hfill
	\begin{minipage}{0.29\textwidth}
		\centering
		\includegraphics[width=1.245\textwidth]{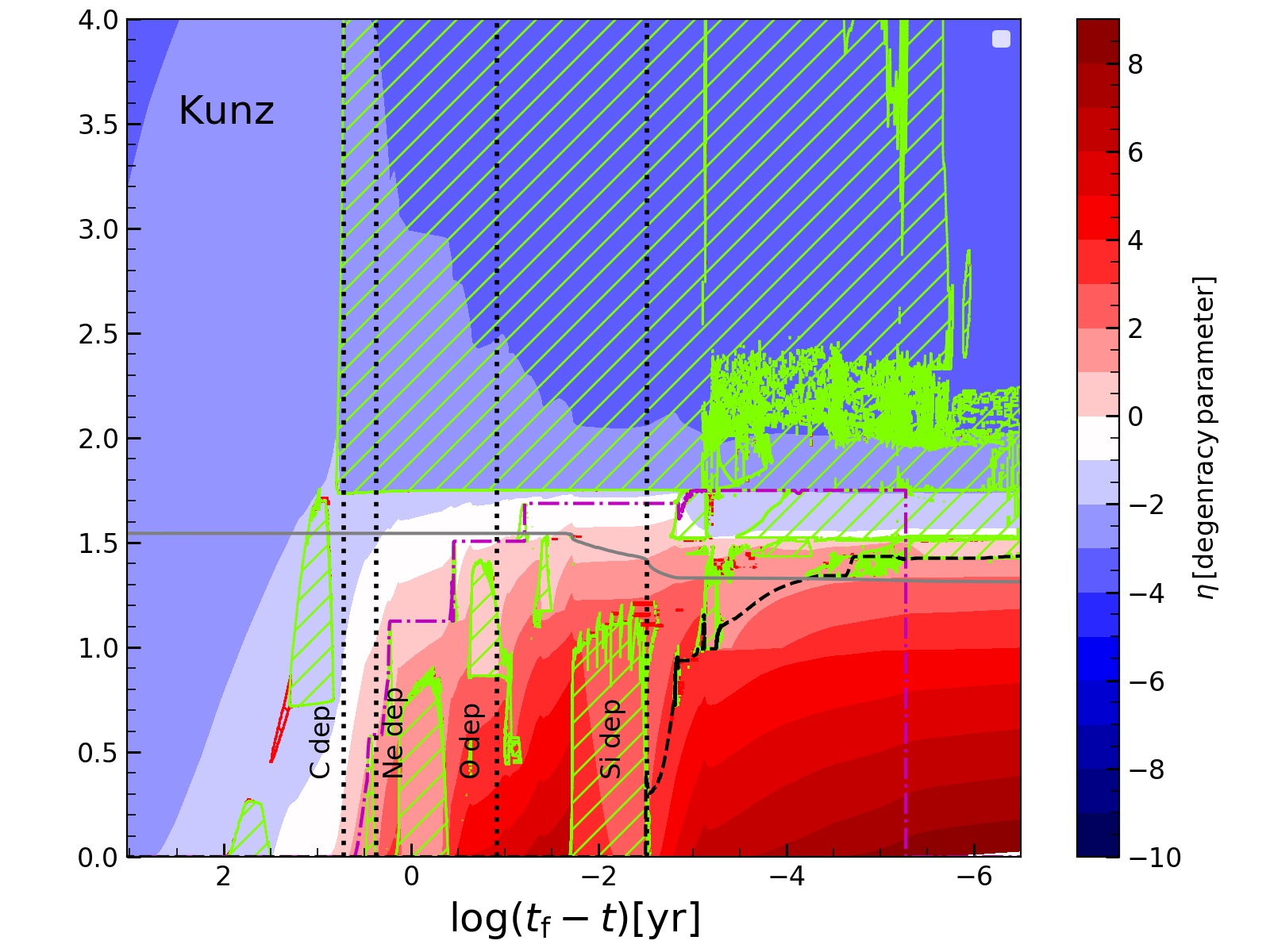}
	\end{minipage}
	\caption{Same as Fig. \ref{fig:reaction_rate_kippen}, but the color bar represents the dimensionless electron degeneracy parameter, $\eta \equiv \mu / k_{\rm B} T$, where $\mu$ is the chemical potential, $k_{\rm B}$ is the Boltzmann constant, and $T$ is the temperature. For $\eta \ll -1$: non-degenerate; $\eta \approx 0$: partially degenerate; $\eta \gg 1$: strongly degenerate. \label{fig:reaction_rate_eta}}
\end{figure}

 The differences in the core structure are also reflected in the evolution of the compactness parameter. Fig. \ref{fig:compactness} shows the evolution of $\xi_{2.5}$ for three different ${}^{12}\mathrm{C}(\alpha, \gamma)^{16}\mathrm{O}$ reaction rates in 14 $\rm M_{\odot}$ He stars from He-ZAMS to the onset of core collapse. During the core He burning, the $\xi_{2.5}$ of all models increases at the same rate, due to the overall contraction of the core, with energy primarily dominated by the $3\alpha$ reaction. Following core He depletion, the core continues to contract until central C ignition is reached. However, after central C ignition, the evolution of $\xi_{2.5}$ in all models begins to diverge, showing significant differences. This divergence is primarily due to the nature of central C burning, the complex interactions between various convective shells, and the influence of electron degeneracy pressure \citep{brown2001formation,sukhbold2014compactness,renzo2017systematic,chieffi2020presupernova,sukhbold2020missing,patton2020towards,laplace2025s}.

Fig. \ref{fig:reaction_rate_kippen} shows the Kippenhahn diagrams of the inner $4.0 \, \rm M_{\odot}$ core structure from C ignition to core collapse for models with different ${}^{12}\mathrm{C}(\alpha, \gamma)^{16}\mathrm{O}$ reaction rates. In models with different ${}^{12}\mathrm{C}(\alpha, \gamma)^{16}\mathrm{O}$ reaction rates, lower reaction rates lead to higher $X_{\rm i}({}^{12}\mathrm{C})$, setting the initial conditions for core C burning. As $X_{\rm i}({}^{12}\mathrm{C})$ increases, the nuclear energy generation rate $\epsilon_{\mathrm{n}}$ exceeds the neutrino energy loss rate $\epsilon_{v}$, triggering the transition of central C-burning from radiative to convective, accompanied by variations in the location and strength of the convective C-burning shells. For instance, in the model using the NACRE rate (see the left panel of Fig. \ref{fig:reaction_rate_kippen}), central C-burning forms radiatively. After its exhaustion, the significant decline in $\epsilon_{\mathrm{n}}$ induces core contraction, which in turn triggers self-regulation of the core structure. This mechanism drives the rapid expansion and cooling of the outer layers, forming the first C-burning convective shell, followed by the second C-burning convective shell. Thus, the location of the burning front, which reflects the extent of core contraction below, corresponds to a mass coordinate of $M_{\rm r} \approx 1.16 \,\rm M_{\odot}$ for the first C-burning convective shell.
In contrast, the models using the CF88 and Kunz reaction rates (see the middle and right panels of Fig. \ref{fig:reaction_rate_kippen}), a small convective C-burning core develops, along with two gradually contracting convective C-burning shells. In these models, The mass coordinates of the burning fronts for the first C-burning convective shell are approximately $0.90\,\rm M_{\odot}$ and $0.71\,\rm M_{\odot}$, respectively. These values are lower than the mass coordinates of the burning front in the NACRE model, indicating that the core is more expanded during this phase. As shown in Fig. \ref{fig:reaction_rate_eta}, the CF88 and Kunz models exhibit lower electron degeneracy in the core during this phase compared to the NACRE model.

At the end of core C burning (triangle in Fig. \ref{fig:compactness}), the NACRE model exhibits a longer C-burning duration but lower $\xi_{2.5}$ compared to the CF88 model. This is due to the larger first C-burning convective shell in the NACRE model, which falls within the range used to evaluate compactness, thereby reducing the $\xi_{2.5}$. Additionally, the larger convective shell transports fresh ${}^{12}\mathrm{C}$ into it, further prolonging the duration of this burning episode.

\subsection{Advanced burning phases}\label{sect:Beyond C-burning phase}

Following the exhaustion of fuel in the first C-burning convective shell, the core contracts further, triggering Ne ignition at the center and forming a convective Ne-burning core (see also Fig. \ref{fig:reaction_rate_kippen}). The energy released from this process decelerates the advancing burning front, which eventually reaches the base of the second C-burning convective shell at the mass coordinates of $2.92 \, \rm M_{\odot}$, $2.10\, \rm M_{\odot}$, and $1.75\, \rm M_{\odot}$ until the end of the evolution, for the NACRE, CF88, and Kunz models, respectively.

After the depletion of core Ne burning (cross in Fig. \ref{fig:compactness}), the growth rate of $\xi_{2.5}$ differs significantly. For the NACRE model, $\xi_{2.5}$ increases significantly overall, because the mass coordinate at the base of the second C-burning convective shell is greater than the mass coordinate at which we evaluate the compactness. However, the evolution of $\xi_{2.5}$ displays two distinct peaks following the depletion of core O-burning (open circle in Fig. \ref{fig:compactness}), with $\xi_{2.5}$ decreasing sharply after each peak. These oscillations in $\xi_{2.5}$ are primarily driven by the Si- and O-burning convective shells \citep{sukhbold2014compactness,renzo2017systematic,chieffi2020presupernova}. The final $\xi_{2.5}$ is 0.44, corresponding to a final C-free core mass of 2.73 $\rm M_{\odot}$ (magenta dash-dotted line in Fig. \ref{fig:reaction_rate_kippen}), and the mass of the final iron core is 1.79 $\rm M_{\odot}$ (black dashed line in Fig. \ref{fig:reaction_rate_kippen}), which is constrained by the C-free core, determined by the location of the last carbon-burning shell \citep{brown2001formation, schneider2021pre, laplace2025s}.
In contrast, for the CF88 and Kunz models, $\xi_{2.5}$ remains largely constant or decreases slightly, due to the mass coordinate at the base of the second C-burning convective shell being lower than the mass coordinate at which compactness is evaluated. Additionally, the oscillations in the subsequent evolution of $\xi_{2.5}$ are also driven by various convective shells. Ultimately, these two models form C-free cores with masses of 2.06 $\rm M_{\odot}$ and 1.75 $\rm M_{\odot}$, and corresponding iron cores with masses of 1.54 $\rm M_{\odot}$ and 1.46 $\rm M_{\odot}$, resulting in lower $\xi_{2.5}$ values of 0.16 and 0.08, respectively (see Table \ref{physical parameters}).

These results demonstrate how the ${}^{12}\mathrm{C}(\alpha,\gamma)^{16}\mathrm{O}$ reaction rate affects the pre-SN core structure. Our analysis identifies three distinct outcomes, as summarized in Table \ref{physical parameters}. The model using the NACRE rate exhibits the highest final compactness, $\xi_{2.5}=0.44$, with $M_{4}=2.15$ and $\mu_{4}=0.13$. According to the criteria of \citet{o2011black} and \citet{ertl2016two}, this implies a likely failed explosion and BH formation, with the BH mass $M_{\text{rm,grav}} \approx 8.79 \, \rm M_{\odot}$ computed using the semianalytic approach of \citet{muller2016simple}.
However, for the models using the CF88 and Kunz rates, both criteria consistently predict a successful explosion, leading to the formation of a NS.

\section{Discussion}\label{sect:Discussion}
We investigated the effects of ${}^{12}\mathrm{C}(\alpha, \gamma)^{16}\mathrm{O}$ reaction rates on the evolution of a 14 $\rm M_{\odot}$ He stars and its pre-SN core structure, using three representative reaction rate combinations.
As shown in Section \ref{sect:Results}, variations in the ${}^{12}\mathrm{C}(\alpha, \gamma)^{16}\mathrm{O}$ reaction rates significantly impact the pre-SN core structure, not only during the core He burning phase but also throughout the subsequent burning phases, particularly the C burning phase. In this phase, higher reaction rates reduce the available fuel for central C burning, leading to the transition from convective to radiative C burning, which is consistent with previous studies \citep[e.g.][]{brown2001formation, Imbriani2001reaction, woosley2002evolution, The2007ApJ, west2013ApJ, sukhbold2020missing, patton2020towards}. 

Additionally, as shown in Fig. \ref{fig:reaction_rate_kippen}, the increase in the reaction rate extends the C-burning convective shells and shifts their locations, a result also observed in earlier studies \citep{brown2001formation, Imbriani2001reaction, woosley2002evolution, Eid2004Evolution, The2007ApJ}. 
However, \cite{Pepper2022MNRAS} found that higher reaction rates increase the core He burning lifetime in low- to intermediate-mass stars. In contrast, we found that reaction rates do not significantly affect the core He burning lifetime but significantly alter the lifetime of the core C burning phase. These differences in burning phases lead to variations in both the iron core mass and the core compactness parameter, which affect its explodability.

In this work, we considered the uncertainties of the ${}^{12}\mathrm{C}(\alpha, \gamma)^{16}\mathrm{O}$ reaction rate, while excluding the effects of uncertainties in the $3\alpha$, ${}^{12}\mathrm{C}+{}^{12}\mathrm{C}$, and ${}^{12}\mathrm{C}+{}^{16}\mathrm{O}$ reaction rates on the core structure.
However, it is important to emphasize that the pre-SN core structure is also highly sensitive to these unconsidered key nuclear reactions, particularly during the advanced phases of evolution \citep{Bennett2012MNRAS, Pignatari2013ApJ, Fang2017PhRvC,deboer2017c}.

Among these reactions, although the $3\alpha$ reaction rates are better constrained, even a 10\% variation plays a significant role \citep{Austin2005NuPhA}. This is because the $3\alpha$ reaction competes with the ${}^{12}\mathrm{C}(\alpha, \gamma)^{16}\mathrm{O}$ reaction in determining the ${}^{12}\mathrm{C}$ abundance, which directly governs the conditions for C burning and ultimately affects both the core structure and the s-process yields \citep{Tur2007ApJ,Tur2010ApJ}.
In addition, the ${}^{12}\mathrm{C}+{}^{12}\mathrm{C}$ fusion reaction, the primary energy source during core C burning, significantly affects the core structure of the star.
According to \cite{Bennett2012MNRAS}, an increased ${}^{12}\mathrm{C}+{}^{12}\mathrm{C}$ reaction rate triggers earlier core C burning at lower temperatures, reducing neutrino losses and extending the burning lifetime. This uncertainty can also significantly affect the nature of core C burning (whether convective or radiative) and the number and locations of convective shell episodes. These changes ultimately impact both the pre-SN core structure and the final fate of stars \citep{The2007ApJ,Bennett2012MNRAS, Pignatari2013ApJ, Chieffi2021ApJ,Monpribat2022A&A, Dumont2024A&A}. Furthermore, the ${}^{12}\mathrm{C}+{}^{16}\mathrm{O}$ reaction rate can further contributes to energy generation during the merger of O- and C-burning shells, leading to a decrease in the core's compactness \citep{Andrassy2020MNRAS,Roberti2025A&A,laplace2025s}. 
The effect of the uncertainties in these reactions is beyond the scope of this work. However, they may still affect the core structure and the final fate of massive stars. In the future, we will further investigate their effect on the pre-SN evolution and explosion mechanism, with a particular focus on explodability.

\section{Conclusion}\label{sect:Conclusion}

In this work, we investigated the impact of uncertainties in the ${}^{12}\mathrm{C}(\alpha,\gamma)^{16}\mathrm{O}$ reaction rate on the evolution of He stars, focusing on their pre-SN core structures by characterizing the iron core mass $M_{\rm Fe}$ and compactness parameter $\xi_{2.5}$. Our conclusions are as follows:
\begin{itemize}
	
\item 

We found that the ${}^{12}\mathrm{C}(\alpha, \gamma)^{16}\mathrm{O}$ reaction rate has varying degrees of effect on the $M_{\rm CO}$ and the $X_{\rm i}({}^{12}\mathrm{C})$ at the end of core He burning, but significantly affects the subsequent core structure. Specifically, higher reaction rates lead to lower $M_{\rm CO}$ and $X_{\rm i}({}^{12}\mathrm{C})$, resulting in the transition of central C burning from the convective to the radiative regime. This transition induces stronger neutrino losses, causing greater core contraction and a larger C-free core, which consequently leads to the formation of a larger iron core and a high value of compactness. These changes indicate a more compact pre-SN structure, which is harder to explode and is more likely to result in a BH remnant.

\item We traced the evolution of compactness using different ${}^{12}\mathrm{C}(\alpha, \gamma)^{16}\mathrm{O}$ reaction rates and found that increasing this reaction rate significantly shifts the location of the last C-burning convective shell. When the location of the last C-burning convective shell exceeds the mass coordinate at which compactness is evaluated, the compactness increases significantly overall; otherwise, it remains essentially unchanged. Furthermore, the Si- and O-burning convective shells induce oscillations in compactness.

\item A variation of the ${}^{12} \mathrm{C}(\alpha, \gamma)^{16} \mathrm{O}$ reaction rates affects the core C burning lifetime, with a lower reaction rate leading to a shorter lifetime.
This triggers earlier central Ne ignition, which slows down the core contraction and prevents the C-burning front from advancing to higher mass coordinates. Ultimately, this leads to smaller iron core masses and lower final compactness.

\end{itemize}

\begin{acknowledgements}
This study is supported by the National Natural Science Foundation of China (Nos 12288102, 12225304, 12090040/12090043, 12273105), the National Key R\&D Program of China (No. 2021YFA1600404), the Youth Innovation Promotion Association of the Chinese Academy of Sciences (No. 2021058), the Yunnan Revitalization Talent Support Program (Yunling Scholar Project and Young Talent Project), the Yunnan Fundamental Research Project (No 202201BC070003, 202401AV070006, and 202201AW070011), and the International Centre of Supernovae, Yunnan Key Laboratory (No. 202302AN360001).
\end{acknowledgements}

\bibliographystyle{raa}
\bibliography{refer}

\end{document}